\documentclass[12pt]{iopart}
\usepackage{graphicx}
\usepackage[dvipsnames,usenames]{xcolor}
\def\br{\begin{eqnarray}}
\def\er{\end{eqnarray}}
\def\be{\begin{equation}}
\def\ee{\end{equation}}

\begin{document}
\title[Photoinduced Quantum Tunneling ]{Photoinduced Quantum Tunneling Model Applied to an Organic Molecule}

\author{E Drigo Filho$^1$, K H P Jubilato$^2$ and R M Ricotta$^3$}

\address{$^{1,2}$Instituto de Bioci\^encias, Letras e Ci\^encias
Exatas, Universidade Estadual Paulista ``J\'ulio de Mesquita Filho", IBILCE-UNESP }

\address{$^3$Faculdade de Tecnologia de S\~ao Paulo, FATEC/SP-CEETEPS\\
Avenida Tiradentes 615 - 01124-060 S\~ao Paulo, SP, Brazil}
\ead{regina@fatecsp.br}
\vspace{10pt}
\begin{indented}
\item[]November 2019
\end{indented}

\begin{abstract}
The paper proposes a photoinduced quantum tunneling model of electron transfer through four quantum square wells potential to simulate the biological process of photosynthesis in bacteria. The problem is mathematically exact with mathematical transcendental equations solved graphically. This simplified model allowed the calculation of the characteristic tunneling times of the process. A comparison is made between the results obtained in the model and the experimental data of the organic molecule Rhodobacter sphaeroides photosynthetic process. 
\end{abstract}

\section{Introduction} 

Quantum tunneling, the problem of barrier penetration, is one of the most interesting and rich phenomena of quantum mechanics. Witnessing them  are the multiple applications in quantum physics, from the most classic examples such as the ammonia molecule, $\alpha$-decay in nuclear physics,  the scanning tunneling microscope and many other effects in solid state physics, as in semiconductors, such as Esaki diode and the Josephson junction,\cite{Merzbacher}, \cite{Razavy}.  Semiconductor devices such as diodes, transistors and integrated circuits can be found in everyday life, on televisions, cars, washing machines,  diminished computers, quantum computers and so on. The tunneling phenomenon is also responsible for the high resolution field emission tunneling microscopy devices and manifests itself in complex theories of measurement fields through the known instantons, \cite{Razavy}. In biological systems quantum tunneling is centered on low temperature electron transport in proteins, \cite{DeVault2}; see \cite{McFadden} for a recent overview.  Quantum biology effects are treated  in recent papers, \cite{Cao}, \cite{Lambert}, \cite{Brookes}.

For many decades the electron transfer has been the subject of much research, \cite{Marcus}, \cite{Gray}-\cite{Xin}. Photosynthesis is one of the phenomena in which electron transfer is carried out with great efficiency, \cite{Mohseni}, occurring in plants, algae, bacteria and cyanobacteria. A process that has generated interest is the electron transfer in cytochrome c \cite{{Hodges}, {Takano}}. In \cite{DeVault1}, DeVault and Chance  suggested that a quantum treatment should be given to the electron transfer of this molecule. Then, experiments showed the temperature independence in electron transfer processes at low temperatures. In these experiments it can be observed that the charge transfer depends on the temperature up to a critical temperature value ($T_c$);  for lower temperatures the transfer remains independent. This independence led to the conclusion that the electron transfer, under these conditions, occurs by electronic quantum tunneling \cite{DeVault2}.

In \cite{Hopfield}, using the Forster's theory \cite{Forster}, Hopfield studied the electron transfer between two fixed sites under two conditions: high and low temperatures. The results were compared with previous results obtained on electron transfer in {\it Chromatium} and in {\it Rhodopseudomonas sphaeroides} bactheria;  the distance dependence in the transference is discussed by a square barrier model. 
 
Currently, studies of systems with high energy conversion efficiency, such as the photosynthetics, have been growing more and more, especially in photovoltaic cell research. One of the approaches used to study these processes involves the structure of multiple quantum wells,  \cite{Vijaya}.

Many photosynthetic reaction centers are known, among them the reaction center of the purple bacterium  {\it Rhodobacter sphaeroides}  that is  organic molecule composed mainly by bacteriochlorophylls, pheophytins, quinones and cytochromes. In bacterial photosynthesis electron transfer occurs in cycles. The process starts with an infrared absorption which allows the electron excitation from the ground state to an excited state. Then, the electron is transferred by charge separation (donor-acceptor) to the cytochrome, which is the protein responsible for electron transport \cite{{Warren},{Wang}}. At the end of the process  the electron returns to its initial state, completing the cycle.

Inspired by all above information, this work proposes a temperature independent quantum mechanical model for the  analysis of a photosynthetic system, based on a specific type of architecture of four asymmetric one-dimensional quantum square wells potential to study  analytically the photoinduced driving electron tunneling in an organic molecule system. The chosen approach aims at a determination of the energy eigenvalues by taking as a starting point  the depth of the potential wells allied with a direction with which the electron is transferred along the wells. The transfer steps occur in pairs, from one well to its neighbor; the asymmetry helps to direct the electron transfers. The energy eigenvalues for the proposed potential can be determined exactly by solving a sequence of asymmetric one-dimensional double well problem,  each one characterized by a two level system, for which resonant states can be determined, \cite{Paulino}. The Rabi formula \cite{Cohen-Tannoudji} can be applied to determine the tunneling time from one well to another. This time will be compared with the ones of the {\it Rhodobacter sphaeroides} purple bacteria.

This work is divided as follows: in Section 2 the bacterial photosynthesis is presented. In Section 3 the one-dimensional model of four asymmetric quantum square wells potential and the solution is described. Section 4 presents the quantitative results obtained for the eigenvalues and time of each transfer. Section 5 contains the conclusions.

\section{Bacterial Photosynthesis} 

As our interest focuses the bacterial photosynthesis, in particular the {\it Rhodobacter sphaeroides} purple bacteria, for which the  photosynthetic reaction center is known, \cite{{Warren},{Wang}}, it is opportune a preliminar description of the biological process. Initially there is the absorption of a photon that causes the system to go from the ground state to an excited state. This initial energy is fundamental for the electron begin to be transferred on the process to its initial state, in a cyclical process. The wavelength absorbed by the chlorophylls in this specific bacterium is 870nm. These chlorophylls that capture the photons are so essential for the photosynthesis that they are known as ``special pair" and symbolized by the letter P. Then the excited electron decays successively passing by the bacteriochlorophylls (B), bacteria pheophytins (H), quinones (Q). The bacteria pheophytins are very similar to the bacteriochlorophylls but do not have magnesium. Quinone is an oxidizing agent.

Two aspects of this process can be highlighted: the energy levels in which the electron is in each step of the transfer and the times involved in the  transfers. In Table 1 it is possible to observe the values of these related parameters in each step of the electron transfer \cite{{Warren},{Wang}}. 
\begin{table}[h]
\centering
\caption{Separation of redox charges at the photosynthetic reaction center \cite{{Warren},{Wang}}, where P symbolizes a special pair of chlorophylls, B is bacteriaclorophyll, H is bacteria pheophytin and Q symbolizes quinones.}
\vspace{0.5cm}
\begin{tabular}{l|c|c}
Step of the Transfer& $\Delta$E (eV) & Time \\ \hline                               
$P-H-Q  \rightarrow {}^*P-B-H-Q$& 0 - 1.40 &  \\ \hline 
${}^*P-B-H-Q \rightarrow P^+-B^--H-Q$ & 1.40 - 1.30 & 3ps   \\ \hline 
$P^+-B^--H-Q  \rightarrow P^+-B-H^--Q$& 1.30 - 1.15  & 1ps  \\ \hline 
$P^+-B-H^--Q \rightarrow  P^+-H-Q^-$ & 1.15 - 0.65  &200ps  \\ \hline 
        
\end{tabular}
\end{table}

 The initial excitation of the electron is clearly observed, resulting in a change in the energy level by the absorption of the photon. Next, we notice the decay of the energy levels until the quinones ($Q^ -$).  The electron transfer to neutralize the original site  ($P^ -$) involves special attention and is not discussed here. Figure 1(a)  shows a scheme of the electron pathway through all stages of the transfer.
\begin{figure} 
\centering
\includegraphics[width=1.0\textwidth]{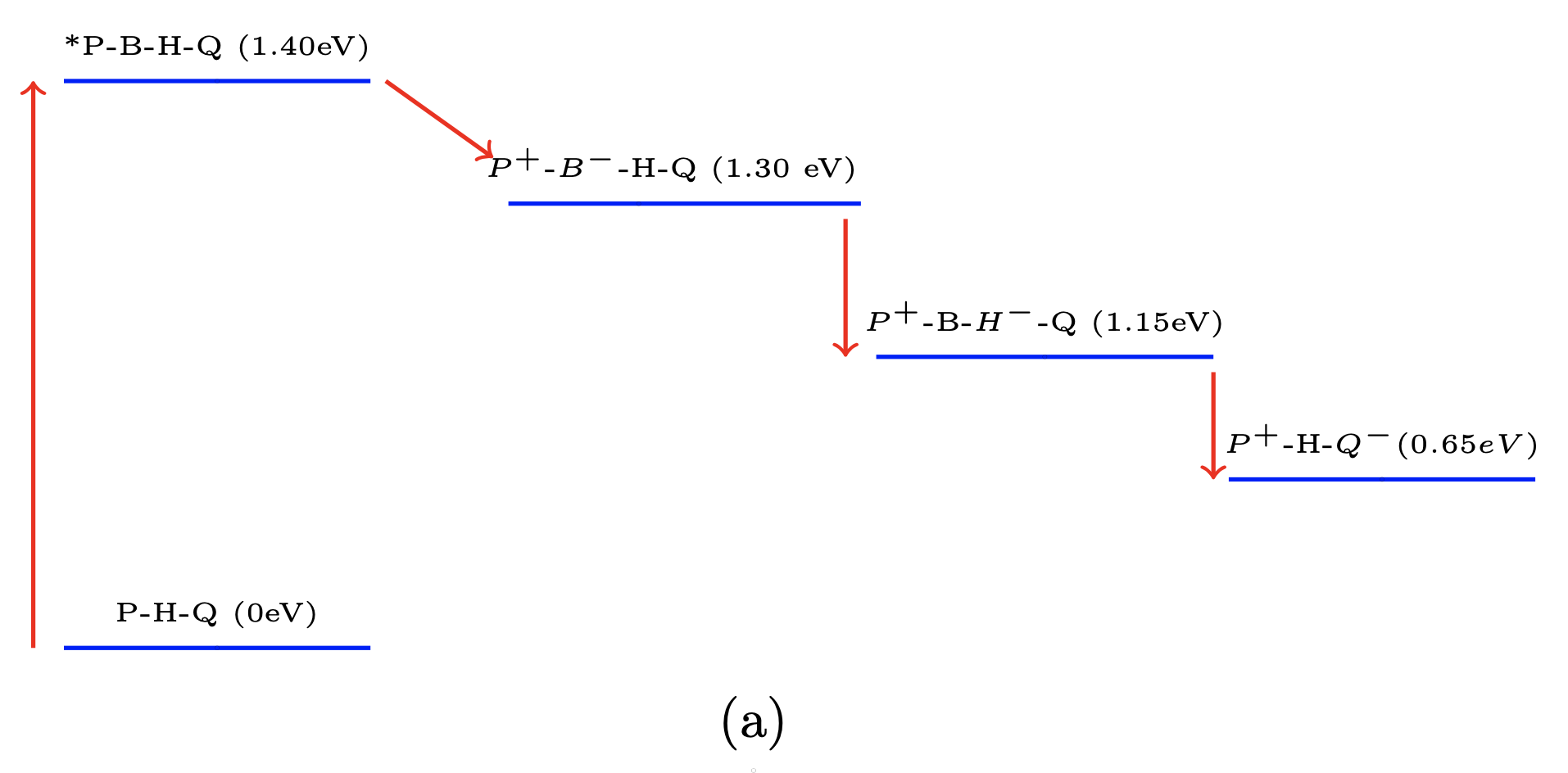}
\includegraphics[width=1.0\textwidth]{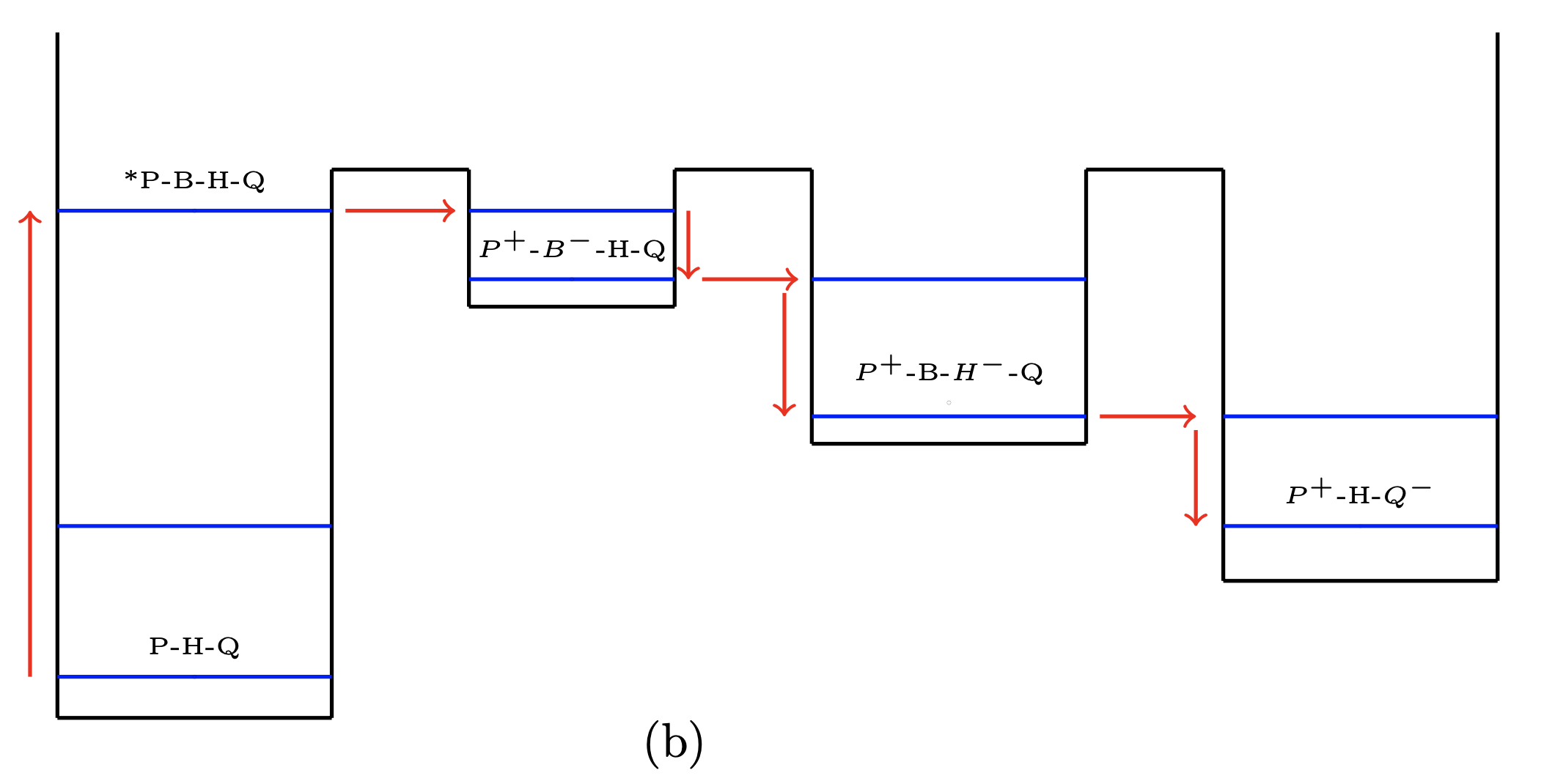}
\caption{\label{label} (a) Stages of electron transfer at the {\it Rhodobacter sphaeroides} reaction center, showing the energies involved and the electron pathway, as described in references \cite{{Warren},{Wang}}. (b) Model of quantum wells for the electron pathway indicated by arrows.}
\end{figure}

\section{The one-dimensional model of four asymmetric square wells potential}

 Since {\it Rhodobacter sphaeroides} photosynthesis is performed in four main steps (transfers),  the proposed model contains four asymmetric one-dimensional quantum square well potentials, as indicated in Figure 1(b). The asymmetry of the wells, with different depths is a necessary feature to direct the electron transfer through all the steps of  the process.  The electron pathway in the model is simulated by an initial absorption, then the electron tunnels and decays through the succession of barriers and wells. The  steps of the biological system are represented by arrows, shown in Figure 1(b). The parameters that need to be fixed are: the width ($a$) of the wells, the distance ($L$) between the center of one well to the center of the neighbour well and value of the potential barriers involved ($V$). All these parameters were fixed aiming for the existence of resonant states in the tunneling process. The initial absorption with  wavelength around  $870nm$ \cite{{Warren},{Wang}} is an essential constraint in the construction of the model, corresponding to a variation of the energy of $1.42 eV$. 
 
The solution of the associated Schr\"{o}dinger equation is analytically exact, thus allowing us to find the eigenfunctions and, with the appropriate boundary conditions, to determine the energy eigenvalues for each region of the potential. As the transfer occurs in pairs of wells, from one well to its neighbor, the solution is obtained by breaking the potential in four sets of asymmetric double wells, where each double well is regarded as the junction of two single ones, and each well confined by an infinite barrier on one side and a finite barrier on the other side, \cite{Paulino}.  Thus the physical problem to be solved involves an asymmetrical bistable well with respect to the depths  $V_0$ and $V_1$ of the two wells involved, where $V_0 <V_1$, as shown in Figure 2.  The transcendental equations obtained together with the wavefunctions for the two regions of interest are given in Table 2.
\begin{figure}
\centering
\includegraphics[width=1.0\textwidth]{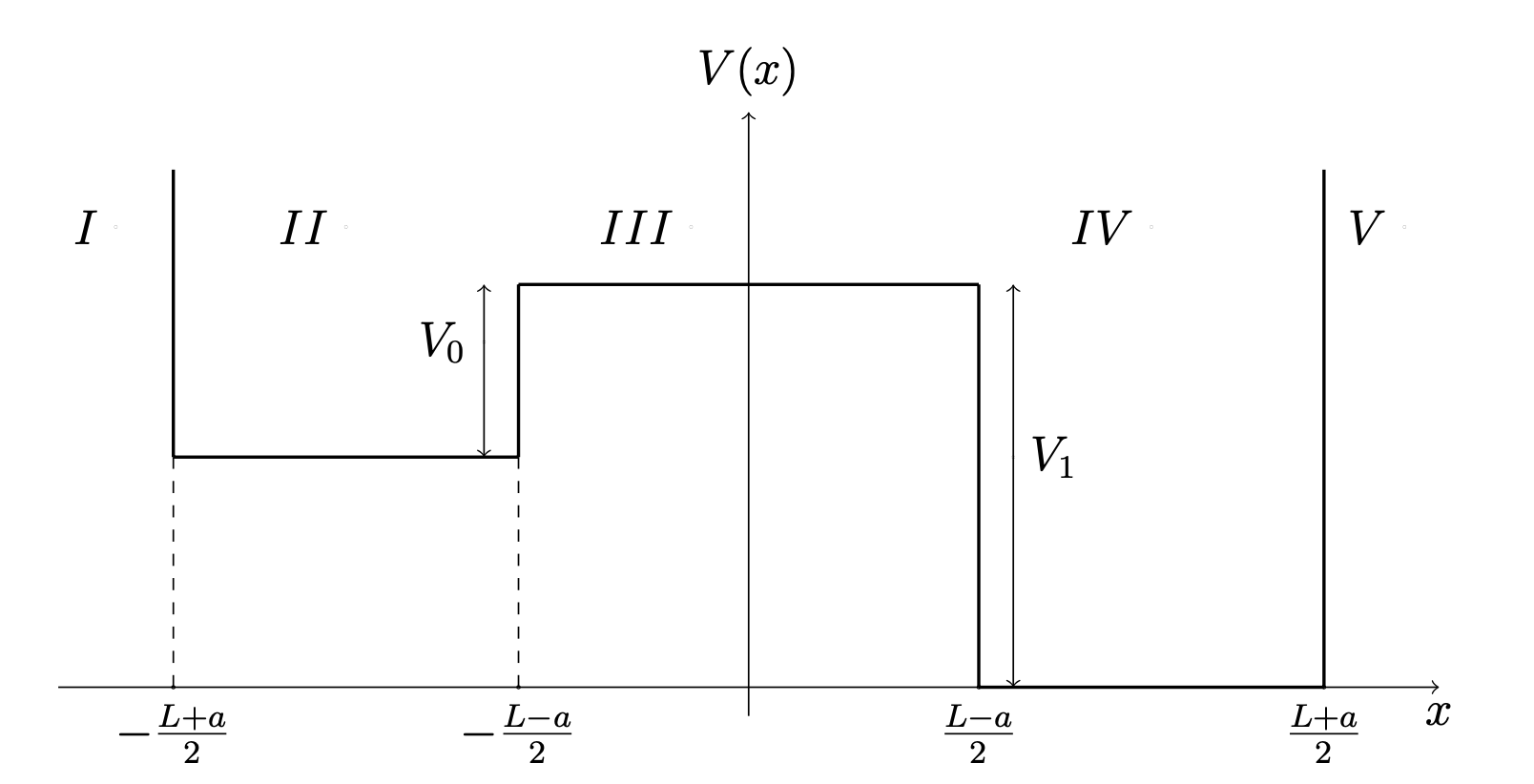}
\caption{\label{label}  Representation of first asymmetric unidimensional double square well potential, showing five different regions, I to V, with $0 < E < V_0$, \cite{Paulino}. }
\end{figure}

Because the wells are asymmetrical, the energy levels of a well relative to its neighbour may or may not match, depending on the depth of the potential. When the levels match these energy levels correspond to resonant states and the system can be approximated by a two-level system. 

\begin{table}
\caption{Wave functions and transcendental equations for the asymmetric one-dimensional double well \cite{Paulino}, where  $L$ is the  distance between the centers of the wells, $a$ is the width of the wells, $m$ is the electron mass, $\hbar= \frac{h}{2\pi}$ where $h$ is Planck's constant, $V_0$ and $V_1$ are the depths of the wells shown in Figure 2 and $E$ is the energy eigenvalue to be determined graphically, with two possibilities: $0 < E < V_0$ and $V_0 < E < V_1 $.}
\begin{tabular}{l}\\
 \hline \hline \\
 Energy: $\;0 < E < V_0\; $\\
Parameters:
$k_1^2 = \frac{2 m}{\hbar^2} ((V_1 - V_0)-E),\;\beta^2 = \frac{2 m}{\hbar^2} (V_1 -E),\;k_2^2 = \frac{2 m}{\hbar^2}E$ \\ 
Wavefunctions:\\                    
Regions I and V: $\;\;\; \Psi_I(x) =\Psi_{V}(x)=0$\\
Region II: $\;\;\;\Psi_{II}(x) = A_1e^{k_1 x}+ A_2e^{-k_1 x}$\\
Region III: $\;\;\;\Psi_{III}(x) = Be^{\beta x}+ Ce^{-\beta x}$\\
Region IV: $\;\;\; \Psi_{IV}(x) = D_1sin(k_2 x)+ D_2cos(k_2 x)$\\ \\
 $f_0(E)=${\large$\;\frac{\left((k_1 - \beta ) e^{k_1(L-a)}+ (k_1 + \beta ) e^{k_1(L+a)}\right)e^{\beta (L-a)}}{((-k_1 + \beta ) e^{k_1(L+a)}- (k_1 + \beta ) e^{k_1(L-a)})}$}; 
$\;\;\;\;\;g_0(E)=${\large$\; \frac{\left(\beta - k_2 cot(k_2 a)\right)e^{-\beta (L-a)}}{\beta + k_2 cot(k_2 a)}$} \\ \\
Transcendental Equation:$\;\;f_0(E)=g_0(E)$\\ \\ \hline \\ 
Energy:$ \;\;\;\;V_0 < E < V_1 $\\ 
Parameters:
$\;\;\;k_1^2 = \frac{2 m}{\hbar^2} (E - (V_1 - V_0)),\;\beta^2 = \frac{2 m}{\hbar^2} (V_1 -E),\;k_2^2 = \frac{2 m}{\hbar^2}E$ \\ 
Wavefunctions:\\
Regions I and V: $\;\;\;\Psi_I(x) =\Psi_{V}(x)=0$\\
Region II: $\;\;\;\Psi_{II}(x) = A_1sin(k_1 x)+ A_2cos(k_1 x)$\\
Region III: $\;\;\;\Psi_{III}(x) = Be^{\beta x}+ Ce^{-\beta x}$\\
Region IV: $\;\;\; \Psi_{IV}(x) = D_1sin(k_2 x)+ D_2cos(k_2 x)$\\ \\
 $f_1(E)=${\large$\;\frac{(\beta + k_1 cot(k_1 a)) e^{\beta(L-a)}} {\beta -k_1 cot(k_1 a)}$};  $\;\;\;\;\;g_1(E)=${\large$\;\frac{(\beta - k_2 cot(k_2 a)) e^{-\beta(L-a)}}{\beta + k_2 cot(k_2 a)}  $}\\ \\
Transcendental Equation:$\;\;f_1(E)=g_1(E)$\\ \\ \hline \hline
\end{tabular}
\end{table}

\section{Results}

After solving the transcendental equations of Table 2 with the constraints, the best value found for the width  of the wells was  $a=43.85\AA$ and the distance $L$ between the wells varies from $60$ to $65\AA$, as shown in Figure 3. These distances are compatible with the distances between the molecules at the  photosynthetic reaction center of the bacterium studied \cite{Warren},  by considering that ($L - a$)  is about $15\AA$. 

The condition that at the beginning of the process there is absorption of light  (with energy of $1.42 eV$) imposes that the first
well of the potential be deeper. Then the electron tunnels and then it decays, which suggests that the electron is transferred in a process of steps ensuring that it passes through all the desired levels. 

To obtain the energy eigenvalues the model is solved in pairs, that is, $V_1$ with $V_2$, $V_2$ with $V_3$, $V_3$ with $V_4$ and $V_4$ with $V_1$, Figure 3.

In this way, solving numerically the transcendental equations for energy between $0$ and $V_0$ , ($0<E<V_0$), and between $V_0$ and $V_1$,  ($V_0<E<V_1$), concerning the solution of the asymmetric double well shown in Table 2, it was found the first pair of wells with the deeper depth being $V_1 = 1.585eV$ and the other with the shallower depth, $V_2 = 0.272eV$. The model is then built in pairs and the junction of these pairs of wells is always done by repeating, in the next pair of wells, one of the values of the potential found. For the second pair of wells the 
potential $V_2$ was used as a constraint, finding then the third well,  $V_3= 0.524eV$ and also the fourth well,  $V_4 = 0.95eV$ (Figure 3), always ensuring the constraints mentioned above.The energy eigenvalues are found graphically, as solutions of the  transcendental equations.

Thus the electron transfer is the following: initially the electron in the ground state ($E_0 = 0.01828eV$) of the first well, $V_1$,  is excited by a photon of wavelength $869.7nm$  ($\sim 870 nm$) to the resonant states $1.445eV$ and $1.460eV$. The length of photon wave is obtained using the value of the energy absorbed, that is, $1.426676eV$. Then, the electron tunnels to the second well, $V_2$, and decays to the ground state $1.329eV$. This state is resonant with a state of the third well, $V_3$, which generates the resonant states of energy $1.329eV$  and $1.335eV$ allowing tunneling. Then the electron decays to the ground state of $V_3$ of energy $1.0785eV$. Further tunneling occurs to the fourth well, $V_4$, with resonant states at  $1.0785eV$ and $1.0787eV$ and the electron decays to the ground state of $V_4$ of energy  $0.6529eV$, as shown in Figure 3. The return to the original state of the system is a complex phenomenon and it is necessary a specific model to describe this process. Thus, this aspect is not analyzed in the present work.
\begin{figure}
\centering
\includegraphics[width=1.0\textwidth]{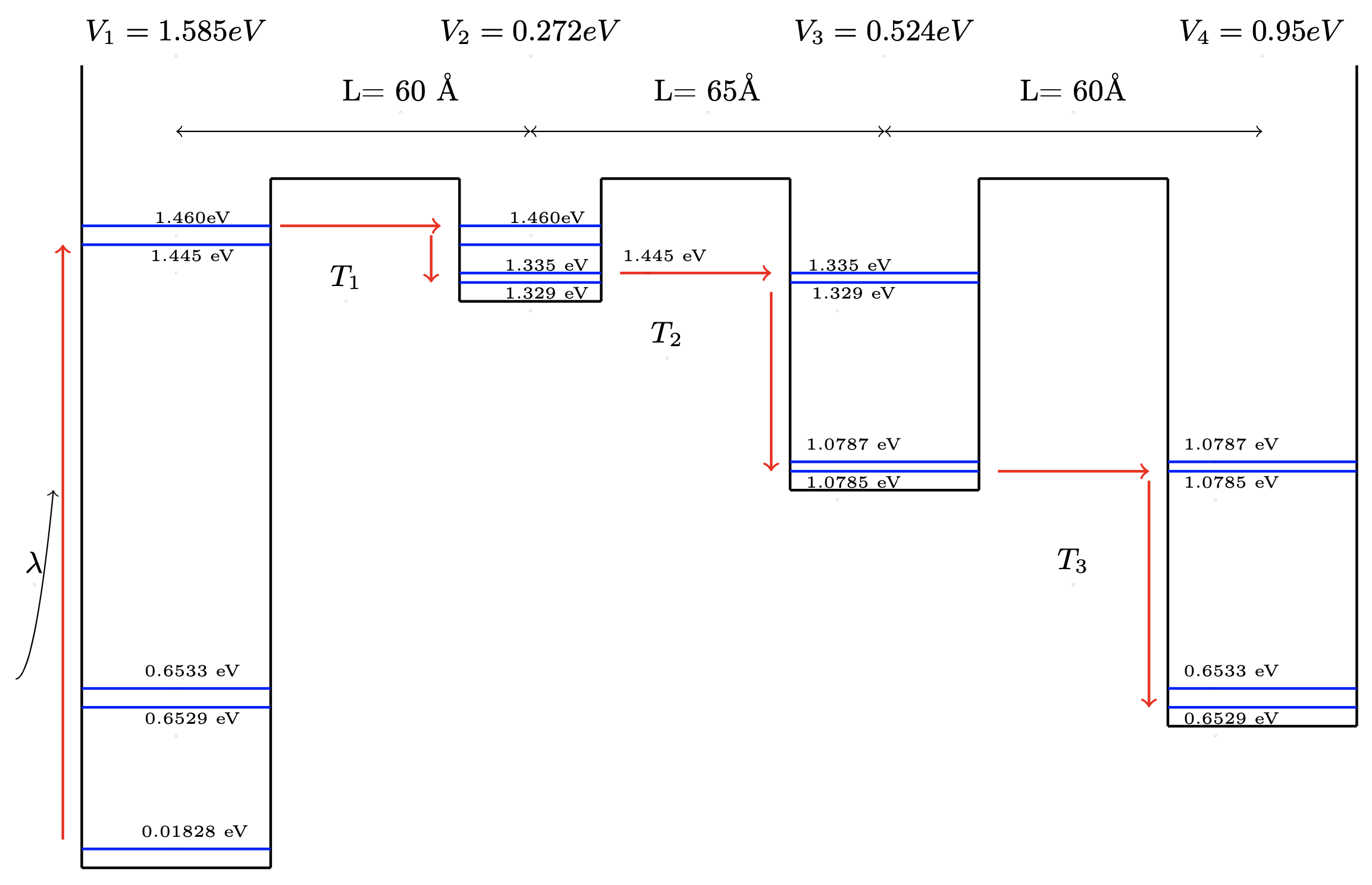}
\caption{\label{label}   Proposed model using four asymmetric quantum wells to describe electron transfer in bacterial photosynthesis. The electron is excited and suffers a succession of tunneling and decays until it returns to the initial state, closing the cycle. $T_1 = 0.14 ps $; $T_2 = 0.35 ps $ and $T_3 = 11 ps$.}
\end{figure}

Figure 4 illustrates the intersection of the functions of the transcendental equation graph for energy within the two wells $V_1$ and $V_2$ with resonant states of energies $1.445eV$ and $1.460 eV$.
\\end{figure}

\begin{figure}
\centering
\includegraphics[width=1.0\textwidth]{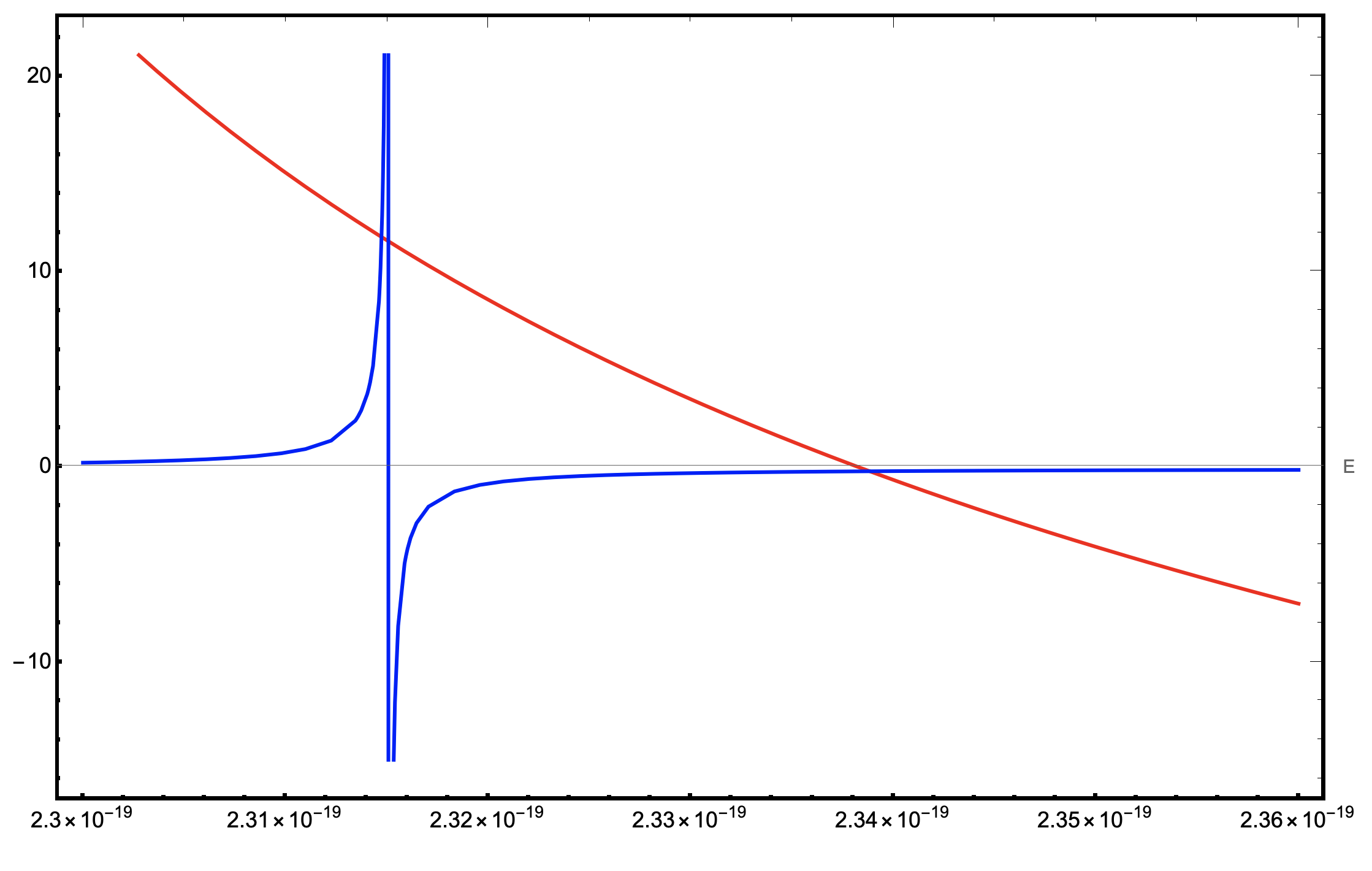}
\caption{\label{label}  Graphic of  transcendental equation, Table 2 for $V_1< E<V_2$, for energies given in joules,    ({\bf \textcolor{red}{-----}} $f_1(E)$,  {\bf \textcolor{blue}{-----}} $ g_1(E)$).  The intersections are at $E = 1.445 eV= 2.3149 .10^{-19}J$ and $E=1.460 eV=2.3389 .10^{-19}J$. Program used: Wolfram  Mathematica.}
\end{figure}
The resonance of states has an important relevance in this work, it reinforces  the probability of electron tunneling through the potential barrier. This probability can be calculated by the formula of Rabi, \cite{Cohen-Tannoudji} presented in the following  equation:
\be
P_{12}(t) = \frac{4|W_{12}|^2}{(E_1 - E_2)^2 + 4|W_{12}|^2}sin^2(\frac{(E_+ - E_-)t}{2 \hbar})
\ee
where $W_{12}$ corresponds to the coupling between the wells involved in the tunneling, $E_1$ and $E_2$
are the original states corresponding to each individual single well and $E_+$ and $E_-$ represent the 
new eigenstates that result from the interaction between the wells.

In case the eigenvalues $E_1$ and $E_2$ are close enough (resonant) the term before the sine function is approximately one and the tunneling time (T) is obtained by considering the maximum value of the sine function, that is, its argument tends to $\frac{(2k+1)\pi}{2}$ for integer values of $k$. Thus 
\be
\label{time} 
T = \frac{(2k+1) \pi \hbar}{E_+ - E_-}\;\;.
\ee

Thus, using  equation (\ref{time}) with $k = 0$, the time for the electron to be transferred from the excited state ($1.460eV$) of well with depth $V_1$ to the ground state (1,329eV) of the well with depth  $V_2$  is $T_1 = 0.14 ps$. Following the same path, the electron is transferred to the well with depth $V_3$ with the time $T_2 = 0.35 ps$ and with the time $T_3 = 11 ps$ to the well with depth $V_4$. These times have an order of magnitude compatible to those obtained experimentally (Table 1).

In order for the process to be efficient it is necessary to compare the tunneling time with the electronic decay time. The decay time can be estimated through the Heisenberg Uncertainty Principle as at least $\Delta t = \frac{\hbar}{2 \Delta E}$, where $\Delta E$ is the difference between the energy levels involved in the transition. The calculations however showed that the tunneling time dominates the process, with at least two orders of magnitude greater than the characteristic transition time.

\section{Conclusions}

Several biological processes, such as photosynthesis, involve electron transfer. In particular bacterial photosynthesis draws attention because it is a driving electron transfer process. 

In this work a photoinduced tunneling model composed by the structure of four asymmetric quantum square wells potential was suggested to simulate the electron transfer within the known reaction center of an organic molecule, the purple bacteria {\it Rhodobacter sphaeroides}. The energy levels and the times required for the steps of the electron pathway were determined exactly. The basic conditions of the biological system were met, including the photon absorption of near $870nm$, the resonant states and the direction of the electron transfer fixed by the asymmetry of the wells, Table 1.

The tunneling and electronic decay times were determined: the electron takes $0,14ps$ to tunnel from the excited state from the first to the second well and decay to the ground  state of the latter. Sequentially, the electron takes $0.35ps$ and $11ps$ to reach the ground state of the third and fourth wells, respectively. The  times obtained analytically by the model are lower than those obtained experimentally, as shown in Table 1 \cite{{Warren},{Wang}}. This discrepancy is attributed to the simplicity of the one-dimensional model studied in this paper which limits a quantitative adjustment in these times. Quantitative results can be improved by increasing the number of physical parameters of the model. For instance,  the individual width (a) of each individual well can be varied, \cite{Oberli}. With such modifications it is expected that the model can be enriched, approaching  the experimental results. Alternatively, improvement of the results may be obtained through the use of an effective potential and parameters adjusted to the description of the physical reality. However, for these cases the solution of the Schrödinger equation no longer has an analytical solution and numerical methods, as presented in   \cite{Degani} and \cite{Souza}, or approximate methods, such as the variational method, \cite{Elso}, need to be employed.

\end{document}